\documentclass[pra, showpacs, twocolumn, floatfix]{revtex4}

\usepackage{graphicx}
\usepackage{amsmath, amsfonts, amssymb, bm}
\usepackage[]{psfrag}

\psfragscanoff
\begin{document}

\title{Polarization operator approach to electron-positron pair production\\ in combined 
laser and Coulomb fields}

\author{A. I. Milstein}
\altaffiliation{Permanent address: Budker Institute of Nuclear Physics, 630090 Novosibirsk, Russia}
\author{C. M\"uller}
\author{K. Z. Hatsagortsyan}
\author{U. D. Jentschura}
\author{C. H. Keitel}
\affiliation{Max-Planck-Institut f\"ur Kernphysik, Saupfercheckweg 1, 69117 Heidelberg, Germany}

\date{\today}

\begin{abstract}
The optical theorem is applied to the process of electron-positron pair creation in the 
superposition of a nuclear Coulomb and a strong laser field. We derive new representations 
for the total production rate as two-fold integrals, both for circular laser polarization 
and for the general case of elliptic polarization, which has not been treated before. 
Our approach allows us to obtain by analytical means the asymptotic behaviour of the pair 
creation rate for various limits of interest. In particular, we consider pair production
by two-photon absorption and show that, close to the energetic threshold of this process,
the rate obeys a power law in the laser frequency with different exponents for linear and
circular laser polarization. With the help of the upcoming x-ray laser sources our results 
could be tested experimentally.
\end{abstract}
 

\maketitle
\section{Introduction}
Studies of electron-positron pair production in strong external fields are an important tool 
to reveal the structure of the QED vacuum. Dynamical pair creation by two Coulomb fields in
relativistic heavy ion collisions has been analysed in great detail, both theoretically and
experimentally (see, e.g., \cite{ion} for a topical review). In recent years, due to the 
sustained progressing of laser technology, 
pair creation in strong laser fields is encountering a growing interest. In several theoretical
papers pair production in the standing wave formed by two counterpropagating laser 
beams has been considered \cite{crossed}. In most cases a laser field strength of order or 
above the critical value $E_c = 1.3\times 10^{16}$ V/cm is required in order to observe the 
process in experiment \cite{ringwald}, which still is by four orders of magnitude larger than 
the highest laser field strengths achievable today \cite{Emax}. More promising in view of its experimental realization is pair production in the combination of a laser and a Coulomb field. 
Here one can exploit the possibility to let a nucleus or a highly charged ion collide at 
large value of the relativistic Lorentz factor 
$\gamma$ with the laser beam. Then, in the nuclear rest frame, the laser's field 
strength $E$ and frequency $\omega$ are enhanced by a factor $\approx 2\gamma$. Observable pair 
creation rates should result for $E\sim E_c$ or $\hbar\omega\sim mc^2$ in this frame
(here, $\hbar$, $c$ and $m$ are Planck's constant, light velocity and electron mass,
respectively). With the help of the upcoming novel laser \cite{XFEL} and accelerator \cite{LHC} 
facilities these conditions can be met. The Large Hadron Collider, that is presently under
construction at CERN (Geneva, Switzerland), will accelerate protons to an energy of 7 TeV 
($\gamma\approx 7000$) \cite{LHC}. When such protons are brought into collision with the 
strongest laser beams available today ($E\approx 10^{-4}E_c$ at $\hbar\omega\sim 1$ eV), 
then the laser field strength in the proton's rest frame approaches or even exceeds the 
critical value. On the other hand, the x-ray free-electron laser (XFEL) facilities presently 
being developed at DESY (Hamburg, Germany) and SLAC (Stanford, USA) are proposed to provide 
spatially coherent and highly brilliant beams of synchrotron radiation
with single-photon energies of up to $8-12$ keV at maximum field strengths of 
$E\sim 10^{-6}E_c$ \cite{XFEL}. When combined with a moderately relativistic ion beam, 
the laser photon energy in the projectile frame can reach the electron's rest energy.
 
We should mention that a few years ago nonlinear electron-positron pair creation in the 
collision of an ultrarelativistic electron beam ($\gamma\approx 10^5$) and an intense 
optical laser pulse was observed at SLAC \cite{SLAC}. In this situation, there are two 
channels for pair production: A direct channel, via laser-photon absorption in the Coulomb 
field of the electron, and an indirect two-step channel (so-called Breit-Wheeler process), 
where first a high-energy photon is generated by Compton backscattering which afterwards 
creates the pair via absorption of laser photons \cite{ritus}. 
In the experiment the contribution of the indirect channel 
was shown to be dominant. In the present paper, however, the indirect production mechanism
will not be addressed. It is of importance only in the case of a light projectile beam. In 
fact, when the incoming electrons were replaced by a beam of heavy particles (e.g., protons) 
the indirect production channel would be strongly suppressed and direct nonlinear pair 
production (by a virtual photon from the nuclear Coulomb field) could be investigated. 
The latter process is the subject of this paper.

The first calculation of pair creation by a Coulomb and a strong laser field is due to
Yakovlev \cite{Yak}. He treated the case of circular laser polarization and derived asymptotic 
formulae for the production rate in certain parameter regimes. Particularly, the total 
cross sections in the perturbative regime ($\xi\ll 1$, with the intensity parameter $\xi$ given 
in Eq.\,(\ref{xi}) below) as well as in the overcritical-field limit ($\xi\gg 1$ and $E\gg E_c$) 
were obtained. Twenty years later, Mittleman \cite{Mit} considered the process in a 
linearly polarized laser wave. His calculation of the leading term for the total cross section 
in the perturbative multiphoton regime ($\xi\ll 1$ and $\hbar\omega\ll mc^2$) led him to the conclusion that, for the laser intensities and frequencies available at that time, this 
cross section was the smallest one on record. Despite of this pessimistic result, 
another twenty years later the great headway in laser technology that was following the invention 
of chirped-pulse amplification \cite{CPA} triggered newly revived interest in the process. 
Dietz and Pr\"obsting \cite{DP} investigated the influence of a highly charged ion of charge 
$Z \sim 100$ within a nonperturbative numerical approach. However, to make their coupled-channel 
calculations feasible they treated the laser field in dipole approximation. While in the earlier 
work the nucleus was always assumed at rest, M\"uller, Voitkiv, and Gr\"un explored in detail the improved opportunities arising from laser-nucleus collisions. In the articles \cite{MVGff1,MVGff2} 
they performed 
numerical calculations for tunneling pair production by an ultrarelativistic nucleus colliding 
with a superintense near-infrared laser beam and for multiphoton pair creation by a moderately 
relativistic ion beam in combination with an intense x-ray laser wave. Moreover, they considered 
the variant of bound-free pair production where the electron is created in a bound state of the 
projectile nucleus \cite{MVGbf}. Free pair creation in the collision of a relativistic nucleus with 
an ultrastrong x-ray laser of circular polarization was also treated by Avetissian, Avetissian, 
Mkrtchian, and Sedrakian in Ref.\,\cite{Ave}, where analytical formulae in the high-intensity limit 
and some numerical results on the energy spectrum of the created particles in the nonperturbative multiphoton regime ($\xi\sim 1$ and $\hbar \omega \sim mc^2 $) are given.

Most of the treatments mentioned above fully account for the influence of the external
laser field by using the Volkov solutions to the Dirac equation as basis states in a 
perturbative calculation of first order with respect to the nuclear Coulomb field.
Within this framework, the fully differential pair creation rate can be expressed as a
Fourier series over the number of absorbed laser photons, the coefficients of which 
are essentially given by ordinary or generalized Bessel functions for circular or linear
laser polarization, respectively. Because of the large photon numbers involved, taking this 
sum is one of the main obstacles in this kind of approach, in particular for a linearly 
polarized laser wave. As a consequence, it is difficult to get useable expressions for
the total production rate.

\begin{figure}[h]
\begin{center}
\includegraphics[width=8cm]{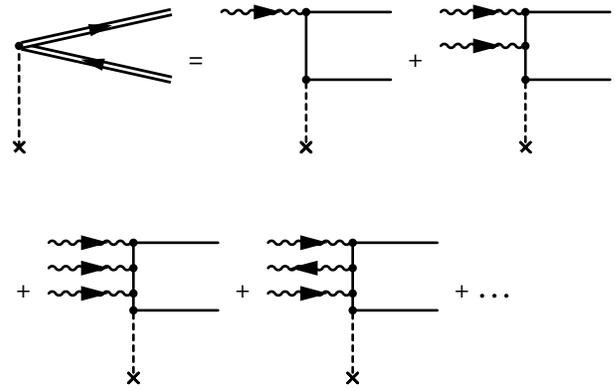}
\caption{\label{laser} Pictorial equation in terms of Feynman diagrams describing 
$e^+e^-$ pair production in combined laser and Coulomb fields. In our approach, the 
Coulomb field is treated within the first order of perturbation theory, while the effect 
of the laser wave is nonperturbatively taken into account to all orders. This is
expressed by the Feynman graph on the left-hand side, where the double lines represent 
the exact lepton wave-functions in the laser field (Volkov states), while the dashed
line stands for their interaction with the Coulomb field. Expanding the Volkov states 
with respect to the lepton-laser coupling results in a perturbation series, some typical 
low-order terms of which are shown on the right-hand side. The wavy lines symbolize the 
laser photons and the arrows indicate whether the respective laser photon is emitted or 
absorbed during the process. The first diagram on the right-hand side, e.g., describes
the leading order of pair production by the net-absorption of one laser photon (with
$\hbar\omega > 2mc^2$). For later reference, we note that the interference between the first 
and the fourth diagram leads to a $\xi^2$-correction to the probability for one-photon pair 
creation [cf. Eq.\,(\ref{delta}) below].} 
\end{center} 
\end{figure}

In the present paper a different approach based on the optical theorem is used to calculate 
the total rate for pair creation in combined laser and Coulomb fields. This approach employs 
the explicit form of the polarization operator of a photon in a laser field found by 
Baier, Milstein, and Strakhovenko \cite{BMS} by means of an operator technique. 
An alternative form of this polarization operator was derived independently by Becker and
Mitter \cite{BM} by means of another method. Both results are in agreement with each other. 
Similar to Refs.\,\cite{Yak,Mit,MVGff1,MVGff2,Ave} the
interaction of the leptons with the laser field is taken into account to all orders
while the effect of the Coulomb field is treated in first order (cf. Fig. 1). The main advantage
of our method is the possibility to derive, independent of the laser's polarization state, 
in an analytical manner compact formulae for the total production rate that only involve
low-dimensional integrals and no additional summations. Based on these representations
one can rather easily find all different kinds of asymptotics. In this way we will confirm 
and extend previous results for a circularly polarized laser wave and give for the first 
time corresponding expressions for elliptic laser polarization. As a main result, we show 
that there are essential differences between the general case of an elliptically polarized
laser field and the special case of circular polarization. Furthermore, because of its interest 
for near-future experiments, particular emphasis is placed on the nonlinear process of pair 
creation by the simultaneous absorption of two photons from the laser wave in a situation,
where the energy of a single photon is not sufficient to produce a pair and where the 
contribution from higher photon orders is neglibly small. We note that, in general, 
contributions stemming from different net-numbers of absorbed laser photons can be 
distinguished via the momentum spectra of the produced particles. In addition, the
strong-field regimes of pair production shall be examined.

The paper is organized as follows. In the next section we describe how the polarization 
operator is related to the process of pair creation, and how it can be used to calculate 
the total pair production rate in the superposition of a Coulomb and a laser field. 
General formulae are given in terms of two-fold integrals for both circular and elliptic 
laser polarization. In the following section we apply these expressions to various 
parameter regions and determine the corresponding asymptotic behaviour of the pair 
production rate. Special attention is paid to the nonlinear process of pair production by 
two-photon absorption in the limit of low laser intensity, for which in particular the 
behaviour close to the energetic threshold is investigated. Moreover, the pair-creation 
cross-section in the high-frequency limit ($\hbar\omega\gg mc^2$) is analysed, where we 
calculate the next-to-leading order correction to one-photon pair production at low
intensity and analyse the limiting case of high laser intensity. Finally, we consider 
the overcritical-field regime as well as the quasiclassical limit of high 
intensity and low frequency, where the rate shows an exponential tunneling behaviour. 
We finish with a conclusion where the significance of our results in view of possible 
experimental investigations is discussed.


\section{Theoretical Framework}

\subsection{The polarization operator and pair production}
The polarization operator $\Pi^{\mu\nu}(k)$ describes the propagation of a photon of
four-momentum $k^\mu$ in a background field (e.g., the QED vacuum), including
self-energy corrections. It is related to the corresponding exact photon propagator
$\mathcal{D}^{\mu\nu}(k)$ via Dyson's equation \cite{LL}
\begin{eqnarray}
\mathcal{D}_{\mu\nu}(k) = D_{\mu\nu}(k) 
  + D_{\mu\sigma}(k){\Pi^{\sigma\lambda}(k)\over 4\pi}\mathcal{D}_{\lambda\nu}(k)
\end{eqnarray}
where $D^{\mu\nu}(k)$ denotes the free photon propagator. In Refs.\,\cite{BMS,BM} the
(properly renormalized) polarization operator was evaluated in the combined fields
of the QED vacuum and a classical electromagnetic plane wave. The interaction with
the vacuum was taken into account to first order in the fine-structure constant
$\alpha=e^2$, with $e$ being the electron charge, while the interaction with the plane-wave 
field was included to all orders. 
In the following, the shape of the electromagnetic wave will be taken as
\begin{eqnarray}
A_\mu^{\rm (L)}(x) = a_{1\mu}\cos(\kappa x)+a_{2\mu}\sin(\kappa x)
\end{eqnarray}
with $\kappa^2 = \kappa a_{1,2} = a_1a_2 = 0$ and the dimensionless intensity parameters 
\begin{eqnarray}
\label{xi}
\xi_{1,2}^{\,2} = -{e^2a_{1,2}^{\,2}\over m^2}.
\end{eqnarray}
Here and henceforth we use relativistic units with $\hbar=c=1$ and write $ab=a^0b^0-{\bf ab}$ 
for the product of two four-vectors. The respective result on the polarization operator of a 
photon in a constant electric field instead of a plane electromagnetic wave can be found 
in Ref.\,\cite{BKS}. Apart from its theoretical 
significance, the knowledge of the polarization operator allows for a number of 
applications. For example, the total probability for $e^+e^-$ pair production by 
a photon of momentum $k$ in an external field (e.g., a plane laser wave) is 
related to the imaginary part of the corresponding polarization operator via \cite{LL}
\begin{eqnarray}
\label{PPph}
W = {\epsilon_\mu\epsilon_\nu^\ast\over k^0}{\rm Im}\Pi^{\mu\nu}(k)
\end{eqnarray}
where $\epsilon_\mu$ denotes the photon's polarization four-vector. In Fig.\,2 we give
an intuitive, pictorial explanation of this formula. 

\begin{figure}[ht]
\begin{center}
\includegraphics[width=2.5cm]{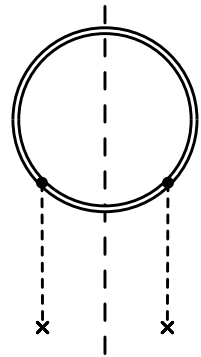}
\caption{\label{pairprod} Lowest order (in $\alpha$) Feynman graph for the polarization 
operator of a photon in an external laser field. In our case the photon is a virtual one 
stemming from a Coulomb field (outer dashed lines). The diagram can be viewed as describing 
the elastic forward scattering of the photon through an intermediate laser-dressed $e^+e^-$ 
state (double line). By performing, as indicated, a cut along the central dashed line, it is 
seen that the graph represents the product of the amplitude for the production of a laser-dressed $e^+e^-$ pair by the virtual photon and its complex conjugate, including a sum over all electron states. Application of the optical theorem (i.e., the unitarity of the scattering matrix) 
therefore relates the imaginary part of the polarization operator with the total probability 
for pair creation, as expressed by Eqs.\,(\ref{PPph}) and (\ref{W}).} 
\end{center} 
\end{figure}

\subsection{Pair creation in combined laser and Coulomb fields}
The calculation in Ref.\,\cite{BMS} was performed for an arbitrary photon momentum $k^\mu$,
including the case $k^2\ne 0$. Therefore, the
polarization operator found there not only applies to the combination 
"photon + laser wave". Instead, the photon may be replaced by any additional external
field $A_\mu^{\rm (ext)}(k)$, e.g. a nuclear Coulomb field. Since a Coulomb field 
can transfer momentum but not energy, here one has $k^\mu=(0,{\bf q})$. To get, at a 
given value of ${\bf q}$, the differential probability (per unit time) for $e^+e^-$
pair creation one has to replace the photon wave-function
$A_\mu^{\rm (ph)}(k)=\sqrt{4\pi/2k^0}\epsilon_\mu$ in Eq.\,(\ref{PPph}) by the 
Fourier transform of the Coulomb field
$A_\mu^{\rm (C)}({\bf q})=(4\pi Ze/{\bf q}^2)\delta_{\mu 0}$. Here, $Z$ denotes the
nuclear charge number. The total production rate then is found by integration over
all possible momenta:
\begin{eqnarray}
\label{W}
\dot W = {(4\pi Ze)^2\over 4\pi}\int {d^3q\over (2\pi)^3}{{\rm Im}\,\Pi^{00}\over q^4}.
\end{eqnarray}
Note that in Eq.\,(\ref{W}) the additional factor of 1/2 compensates for the 
double-counting of the contributions from the momenta ${\bf q}$ and $-{\bf q}$.
For the case of a constant homogeneous electromagnetic field the relation
(\ref{W}) was used in Ref.\,\cite{BKS2}, where the probability for pair production 
was found within the quasiclassical approximation for the polarization operator.

According to the results of Ref.\,\cite{BMS}, the required component of the 
polarization operator in Eq.\,(\ref{W}) can be written as
\begin{eqnarray}
\label{Pi00}
&\Pi^{00}& = -{\alpha\over\pi}m^2 \int_0^\infty {d\rho\over\rho} \int_{-1}^{+1}dv \nonumber\\
& &\!\!\!\!\! \times\exp\left\{ {-2i\rho\over |\lambda|(1-v^2)}
\left[ 1 + {q^2(1-v^2)\over 4m^2} + A(\xi_1^2+\xi_2^2)\right] \right\} \nonumber\\
& &\!\!\!\!\! \times{1\over\cos^2\theta}
\left[ \sin^2\theta\, (d_3 \cos^2\phi + d_4 \sin^2\phi) - d_5 \right],
\end{eqnarray}
where we introduced the angles $\theta = \angle({\bf q},\mbox{\boldmath$\kappa$})$ and 
$\phi = \angle({\bf q},{\bf a}_1)$, and write $\lambda = -\omega q\cos\theta/ 2m^2$.
The coefficients $d_j$ ($j=3,4,5$) in Eq.\,(\ref{Pi00}) are given by Eq.\,(2.32) in 
Ref.\,\cite{BMS}; they read
\begin{eqnarray}
\label{coeff}
d_3 &=& \left( A_1\xi_1^2-\sin^2\rho{\xi_1^2-\xi_2^2\over 1-v^2}\right) [J_0(\zeta)+iJ_1(\zeta)]
\nonumber\\
& & +\xi_1^2\sin^2\rho{1+v^2\over 1-v^2}J_0(\zeta) \nonumber\\
& & +{1\over 4}\left[{q^2\over m^2}-i{|\lambda|(1-v^2)\over\rho}\right][J_0(\zeta)-{\rm e}^{iy}]
\nonumber\\
d_4 &=& d_1(\xi_1 \leftrightarrow \xi_2)\nonumber\\
d_5 &=& {q^2\over 4m^2}(1-v^2)[J_0(\zeta)-{\rm e}^{iy}]
\end{eqnarray}
with the abbreviations
\begin{eqnarray}
A &=& {1\over 2}\left( 1-{\sin^2\rho\over\rho^2} \right),\ \ 
A_0 = {1\over 2}\left( {\sin^2\rho\over\rho^2}-{\sin 2\rho\over 2\rho} \right), \nonumber\\
A_1 &=& A + 2A_0\,, \nonumber\\
\zeta &=& {2\rho A_0 (\xi_1^2-\xi_2^2)\over |\lambda|(1-v^2)}\,,\ \ 
y = {2\rho A (\xi_1^2+\xi_2^2)\over |\lambda|(1-v^2)}.
\end{eqnarray}

For simplicity, we first consider the case of a circularly polarized laser wave 
($\xi_1^2 = \xi_2^2 \equiv \xi^2$). Then the coefficients in Eq.\,(\ref{Pi00}) 
can be simplified to read \cite{BMS}
\begin{eqnarray}
d_{3,4} &=& \xi^2\sin^2\rho{1+v^2\over 1-v^2}
           -{1\over 2}\left[ 1-{q^2\over 4m^2}(1+v^2) \right](1-{\rm e}^{iy})\nonumber\\
d_5 &=& {q^2\over 4m^2}(1-v^2)(1-{\rm e}^{iy})
\end{eqnarray}
with
$$ y = {2\xi^2\rho\over |\lambda|(1-v^2)}\left(1-{\sin^2\rho\over\rho^2}\right).$$
Employing spherical coordinates such that $d^3q = q^2dq\, d\cos\theta\, d\phi$, the 
integration over $\phi$ in Eq.\,(\ref{W}) is trivial and the total rate becomes
\begin{eqnarray}
\dot W &=& -{2\over\pi^2}(Z\alpha)^2m\, {\rm Im} \int_0^1 dv
\int_0^\infty {dQ\over Q^2} \int_0^1 {dt\over t^2} \int_0^\infty {d\rho\over\rho} \nonumber\\
& & \times\,{\rm exp}\left[-i{a\rho\over Qt(1-v^2)}[1+Q^2(1-v^2)]\right] \nonumber\\
& & \times \bigg\{ (1-t^2)\xi^2\sin^2\rho {1+v^2\over 1-v^2} {\rm e}^{-iy} 
+ \bigg[ Q^2(1-v^2) \nonumber\\
& &  + {1-t^2\over 2}\left[ 1-Q^2(1+v^2) \right] \bigg](1-{\rm e}^{-iy}) \bigg\},
\end{eqnarray}
where we have introduced the dimensionless quantities $Q=q/2m$, $t=\cos\theta$, and $a=2m/\omega$.
Going over to the new variables $Q\to Q/\sqrt{1-v^2}$, $t\to t/\sqrt{1-v^2}$ considerably 
simplifies the $v$-dependence such that also this integral can be taken:
\begin{eqnarray}
\label{circ1}
\dot W &=& -{2\over\pi^2}(Z\alpha)^2m\, {\rm Im} 
\int_0^\infty {dQ\over Q^2} \int_0^1 dt \int_0^\infty {d\rho\over\rho} \nonumber\\
& &\!\!\!\! \times\,{\rm exp}\left[-i{a\rho\over Qt}(1+Q^2)\right] 
\bigg\{ \xi^2\sin^2\rho \nonumber\\
& &\!\!\!\! \times\left[{2\over 3}(2+t^2){\sqrt{1-t^2}\over t^2}-
\ln\left({1+\sqrt{1-t^2}\over 1-\sqrt{1-t^2}}\right)\right] {\rm e}^{-iy}\nonumber\\
& &\!\!\!\! +\left[{1\over 3}{(1-t^2)^{3/2}\over t^2}+{Q^2\over 2}
\ln\left({1+\sqrt{1-t^2}\over 1-\sqrt{1-t^2}}\right)\right](1-{\rm e}^{-iy}) \bigg\} \nonumber\\
\
\end{eqnarray}
where now $y=(a\rho\xi^2/Qt)(1-\sin^2\rho/\rho^2)$.
Alternatively, the $Q$-integration can be expressed by Neumann functions $Y_n$:
\begin{eqnarray}
\label{circ2}
\dot W &=& -{(Z\alpha)^2\over\pi}m \int_0^1 dt \int_0^\infty {d\rho\over\rho} \bigg\{
{2\xi^2\sin^2\rho\over z} Y_1(\beta z) \nonumber\\
& & \times\left[{2\over 3}(2+t^2){\sqrt{1-t^2}\over t^2}-
\ln\left({1+\sqrt{1-t^2}\over 1-\sqrt{1-t^2}}\right)\right] \nonumber\\
& & + {2\over 3}\left[ Y_1(\beta)-{1\over z}Y_1(\beta z)\right]{(1-t^2)^{3/2}\over t^2} \nonumber\\
& & + \Big[ Y_1(\beta)- z Y_1(\beta z)\Big]
\ln\left({1+\sqrt{1-t^2}\over 1-\sqrt{1-t^2}}\right) \bigg\}
\end{eqnarray}
with $\beta = 2a\rho/t$ and $z=[1+\xi^2(1-\sin^2\rho/\rho^2)]^{1/2}$. Equation (\ref{circ2}) 
is a very compact, general expression for the total pair production rate in a circularly 
polarized laser field. For applications, both Eqs.\,(\ref{circ1}) and (\ref{circ2}) will 
prove to be useful.

In the general case of an elliptically polarized laser wave the integrations over $\phi$ 
and $v$ can be taken in the same manner as before yielding
\begin{eqnarray}
\label{gen1}
\dot W &=& -{(Z\alpha)^2\over\pi^2}m\, {\rm Im} 
\int_0^\infty {dQ\over Q^2} \int_0^1 {dt\over t^2} \int_0^\infty {d\rho\over\rho} \nonumber\\
& &\!\!\!\! \times\,{\rm exp}\left[-i{a\rho\over Qt}(1+Q^2)\right] {\rm e}^{-iy} \nonumber\\
& &\!\!\!\! \times\Bigg\{ \left(\xi_1^2+\xi_2^2\right)J_0(\zeta)
            \Bigg[ {2\over 3}A_1(1-t^2)^{3/2} \nonumber\\
& &\!\!\!\!  +\sin^2\rho \Bigg({2\over 3}(2+t^2)\sqrt{1-t^2}
             - t^2 \ln\left({1+\sqrt{1-t^2}\over 1-\sqrt{1-t^2}}\right)\Bigg) \Bigg] \nonumber\\
& &\!\!\!\! +i\left(\xi_1^2-\xi_2^2\right)J_1(\zeta)
\Bigg[ {2\over 3}A_1(1-t^2)^{3/2} \nonumber\\ 
& &\!\!\!\! -\sin^2\rho \left(2\sqrt{1-t^2} 
            - t^2 \ln\left({1+\sqrt{1-t^2}\over 1-\sqrt{1-t^2}}\right)\right) \Bigg]\nonumber\\
& &\!\!\!\! + \left[J_0(\zeta)-{\rm e}^{iy} \right] 
\Bigg[{2\over 3}(1-t^2)^{3/2}\left(Q^2-i{Qt\over a\rho}\right) \nonumber\\
& &\!\!\!\! -Q^2t^2\ln\left({1+\sqrt{1-t^2}\over 1-\sqrt{1-t^2}}\right)\ \Bigg] \Bigg\}.
\end{eqnarray}
The integral over $Q$ can again be taken analytically: 
\begin{eqnarray}
\label{gen2}
\dot W &=& {(Z\alpha)^2\over 4\pi}\omega  
\int_0^1 {dt\over t} \int_0^\infty {d\rho\over\rho^2} 
\Bigg\{ {\xi_1^2+\xi_2^2\over \sqrt{g_0^2-g_1^2}} \nonumber\\
& &\!\!\!\! \times\left[ \varphi_- J_1(\varphi_-)Y_0(\varphi_+) 
                       - \varphi_+ J_0(\varphi_-)Y_1(\varphi_+) \right] \nonumber\\
& &\!\!\!\! \times \Bigg[ {2\over 3}A_1(1-t^2)^{3/2} \nonumber\\
& &\!\!\!\!  +\sin^2\rho \Bigg({2\over 3}(2+t^2)\sqrt{1-t^2}
             - t^2 \ln\left({1+\sqrt{1-t^2}\over 1-\sqrt{1-t^2}}\right)\Bigg) \Bigg] \nonumber\\
& &\!\!\!\! - {\xi_1^2-\xi_2^2\over \sqrt{g_0^2-g_1^2}}
            \left[ \varphi_- J_1'(\varphi_-)Y_1(\varphi_+) 
                 - \varphi_+ J_1(\varphi_-)Y_1'(\varphi_+) \right] \nonumber\\
& &\!\!\!\! \times {\rm sign}(g_1) \Bigg[ {2\over 3}A_1(1-t^2)^{3/2} \nonumber\\ 
& &\!\!\!\! -\sin^2\rho \left(2\sqrt{1-t^2}-t^2
\ln\left({1+\sqrt{1-t^2}\over 1-\sqrt{1-t^2}}\right)\right) \Bigg]\nonumber\\
& &\!\!\!\! - \left[ \varphi_- J_1(\varphi_-)Y_0(\varphi_+) 
                   + \varphi_+ J_0(\varphi_-)Y_1(\varphi_+) - \beta Y_1(\beta)\right] \nonumber\\
& &\!\!\!\! \times\Bigg[{2\over 3}(1-t^2)^{3/2} - t^2
\ln\left({1+\sqrt{1-t^2}\over 1-\sqrt{1-t^2}}\right)\Bigg] \nonumber\\
& &\!\!\!\! -{4\over 3} \left[ J_0(\varphi_-)Y_0(\varphi_+) - Y_0(\beta) \right] (1-t^2)^{3/2} \Bigg\}
\end{eqnarray}
with $g_0 = 1 + (\xi_1^2+\xi_2^2)A$, $g_1 = (\xi_1^2-\xi_2^2)A_0$, and
$\varphi_\pm = \left( g_0\pm\sqrt{g_0^2-g_1^2}\right)^{1/2}\!\beta/\sqrt{2}$.

The total production rate can be converted into a cross section according to the
relation
\begin{eqnarray}
\label{flux}
\sigma = {\dot W\over j}\,,\ \ j = {\omega m^2\over 8\pi\alpha}(\xi_1^2+\xi_2^2)
\end{eqnarray}
with $j$ being the photon flux.

\section{Special Cases}
\subsection{One-photon limit}
Let us consider the case of small intensity ($\xi\ll 1$) and frequency $\omega > 2m$.
The leading term in the $\xi^2$-expansion corresponds to pair creation by a single 
photon in a Coulomb field. Under these circumstances the total production probability 
is independent of the laser's polarization state and it is convenient to derive
the asymptotic behaviour by using the formula (\ref{circ1}) for circular laser polarization.
Assuming $\xi\ll 1$, we can expand the exponentials to lowest order with respect to $y$. 
Afterwards we perform the integration over $\rho$ with the result
\begin{eqnarray}
\label{BHtQ}
\dot W  &=& {(Z\alpha)^2\over 2\pi}m\xi^2 \int_0^\infty {dQ\over Q^2} \int_0^1 dt\
\vartheta(2-\gamma) \nonumber\\
& & \times \bigg\{ {2\over 3}(2+t^2){\sqrt{1-t^2}\over t^2}-
\ln\left({1+\sqrt{1-t^2}\over 1-\sqrt{1-t^2}}\right) \nonumber\\
& & + {a\over Qt}\left[{2\over 3}{(1-t^2)^{3/2}\over t^2}+Q^2
\ln\left({1+\sqrt{1-t^2}\over 1-\sqrt{1-t^2}}\right)\right]
\nonumber\\
& & \times \left(1-{\gamma\over 2}\right) \bigg\},
\end{eqnarray}
where $\gamma = a(1+Q^2)/Qt$ and $\vartheta(x)$ is the step function, which restricts the 
integration to the kinematically allowed region $a\le t\le 1$ and $Q_-\le Q\le Q_+$ with 
$Q_\pm = t/a \pm \sqrt{(t/a)^2-1}$. Integrating over $Q$ gives: 
\begin{eqnarray}
\label{BHt}
\dot W &=& {(Z\alpha)^2\over 2\pi}m\xi^2 \int_a^1 dt\ \Bigg\{ 
{2\over 3}\sqrt{{t^2\over a^2}-1}{\sqrt{1-t^2}\over t^2} \nonumber\\
& & \times \left[ {4\over 3}t^2 + {14\over 3} + {2\over 3}a^2 - {2\over 3}{a^2\over t^2} \right] \nonumber\\
& & - 2\left( 1+{a^2\over t^2}\right)\sqrt{{t^2\over a^2}-1}
\ln\left({1+\sqrt{1-t^2}\over 1-\sqrt{1-t^2}}\right) \nonumber\\
& & + {2a\over t}\ln\left({1+\sqrt{1-t^2}\over 1-\sqrt{1-t^2}}\right)
\ln\left({{t\over a}+\sqrt{{t^2\over a^2}-1}}\right) \Bigg\}. \nonumber\\
\ 
\end{eqnarray}
The corresponding expression for the cross section [see Eq.\,(\ref{flux})] agrees with 
the well-known result of Bethe and Heitler \cite{BH}.
Accordingly, in the ultrarelativistic limit ($\omega\gg m$) we find
\begin{eqnarray}
\label{BH}
\sigma = {28\over 9}Z^2\alpha\, r_e^2 
         \left[ \ln\left({2\omega\over m}\right) - {109\over 42} \right]
\end{eqnarray}
with the classical electron radius $r_e = \alpha/m$, and in the nonrelativistic limit 
($\omega-2m\ll m$)
\begin{eqnarray}
\sigma = {\pi\over 12}Z^2\alpha\, r_e^2\left({\omega-2m\over m}\right)^3.
\end{eqnarray}

\subsection{Two-photon pair creation}
It is interesting to consider the cross section for two-photon pair production at
$\xi\ll 1$, which represents a nonlinear strong-field process of lowest possible order 
in the number of
absorbed photons. We assume that $m < \omega < 2m$, which excludes the possibilty of
one-photon pair creation. Going one step further in the $y$-expansion of Eq.\,(\ref{circ1})
and performing the $\rho$-integration, we obtain for the circular polarization case
\begin{eqnarray}
\dot W &=& {(Z\alpha)^2\over 2\pi} m\xi^4 \int_0^\infty {dQ\over Q^2} \int_0^1 dt\
\vartheta(4-\gamma) \nonumber\\
& & \!\!\!\!\!\!\!\! \times\Bigg\{ \left(1-{\gamma\over 4}\right){a\over Qt}
\left[{2\over 3}(2+t^2){\sqrt{1-t^2}\over t^2} \right. \nonumber\\
& & \!\!\!\!\!\!\!\! \left. -\ln\left({1+\sqrt{1-t^2}\over 1-\sqrt{1-t^2}}\right)\right]
+ {1\over 2}\left(1-{\gamma\over 4}\right)^2{a^2\over Q^2t^2} \nonumber\\
& & \!\!\!\!\!\!\!\! \times \left[{2\over 3}{(1-t^2)^{3/2}\over t^2}+Q^2
\ln\!\left({1+\sqrt{1-t^2}\over 1-\sqrt{1-t^2}}\right)\right]\Bigg\}.
\end{eqnarray}
Proceeding as before, we get
\begin{eqnarray}
\label{2ph}
\dot W &=& {(Z\alpha)^2\over 2\pi}m\xi^4 \int_{m/\omega}^1 dt\,
\Bigg\{ {(4t^2-a^2)^{3/2}\over 3at^2} \nonumber\\
& & \times \left[{2\over 3}(2+t^2){\sqrt{1-t^2}\over t^2}-
\ln\left({1+\sqrt{1-t^2}\over 1-\sqrt{1-t^2}}\right)\right] \nonumber\\
& & + {(4t^2-a^2)^{5/2}\over 45at^4}{(1-t^2)^{3/2}\over t^2} + {a\over 6t^4} \nonumber\\ 
& & \times\ln\left({1+\sqrt{1-t^2}\over 1-\sqrt{1-t^2}}\right) \Bigg[ (2t^2+a^2)\sqrt{4t^2-a^2} \nonumber\\
& &  -3a^2t\,\ln\!\left({2t\over a}+\sqrt{{4t^2\over a^2}-1}\right)\Bigg]\Bigg\}.
\end{eqnarray}
After conversion into a cross section, Eq.\,(\ref{2ph}) reads
\begin{eqnarray}
\label{sigmacirc}
\sigma = \alpha Z^2\xi^2 r_e^2 F\left({\omega\over m}\right)
\end{eqnarray}
with 
\begin{eqnarray}
\label{F}
F(x) &=& {8\over 3x} \int_{1/x}^1 dt\, \bigg\{ {[(xt)^2-1]^{3/2}\over (xt)^2} \nonumber\\
& & \times \left[{2\over 3}(2+t^2){\sqrt{1-t^2}\over t^2}-
\ln\left({1+\sqrt{1-t^2}\over 1-\sqrt{1-t^2}}\right)\right] 
\nonumber\\
& & + {4\over 15}{[(xt)^2-1]^{5/2}\over (xt)^4}{(1-t^2)^{3/2}\over t^2} + {1\over (xt)^4} \nonumber\\
& & \times\ln\left({1+\sqrt{1-t^2}\over 1-\sqrt{1-t^2}}\right) \Big[ \left(2+(xt)^2\right)\sqrt{(xt)^2-1}
 \nonumber\\
& &
-3xt\ln\left( xt+\sqrt{(xt)^2-1}\right) \Big] \bigg\}.
\end{eqnarray}
In the limit $0 < x-1\ll 1$, Eq.\,(\ref{F}) becomes
\begin{eqnarray}
\label{Flim}
F(x) &=& {128\over 45} (x-1)^4 \int_0^1 ds\,\sqrt{s}(1-s)^{3/2}(6-s) \nonumber\\
     &=& \pi (x-1)^4.
\end{eqnarray}
Hence, the threshold behaviour of the cross section for two-photon pair production in 
a circularly polarized laser field reads
\begin{eqnarray}
\label{2phlimit}
\sigma = \pi \alpha Z^2\xi^2 r_e^2 \left(\omega-m\over m\right)^4.
\end{eqnarray}
We note that, based on the results of their numerical calculations, the authors of Ref.\,\cite{MVGff2}
found that the rate for two-photon pair creation within the whole frequency range $m < \omega < 2m$
approximately scales as $(\omega-m)^3$. However, Eq.\,(\ref{2phlimit}) now shows that the exact
scaling in the threshold region ($m\lesssim \omega$ and $\omega-m \ll 2m$) is given by 
$(\omega-m)^4$. 
Further, in Ref.\,\cite{MVGff2} a total production rate of $190$ sec$^{-1}$ (in the 
nuclear rest frame) was reported for $Z=1$, $\omega = 900$ keV, and $\xi=7.5\times 10^{-4}$. 
Applying Eq.\,(\ref{2ph}) to these parameters, perfectly reproduces this result.

In the general case of elliptic polarization, the corresponding asymptotic cross section reads
\begin{eqnarray}
\label{sigmagen}
\sigma = \alpha Z^2 {\xi_1^2+\xi_2^2\over 2}r_e^2\,\mathcal{F}\!\left({\omega\over m},\mu\right)
\end{eqnarray}
where $\mathcal{F}(x,\mu)=F(x)+\mu G(x)$ with the function $F(x)$ from Eq.\,(\ref{F}) and
\begin{eqnarray}
G(x) &=& {1\over 3x^2}\int_{1/x}^1 {dt\over t^3}
\bigg\{ {2\over 5}{(1-t^2)^{3/2}\over (xt)^3} \nonumber\\
& & \times \Bigg[ \left((xt)^4-7(xt)^2+16\right)\sqrt{(xt)^2-1} \nonumber\\
& & -10xt\ln\left(xt+\sqrt{(xt)^2-1}\right)\Bigg] \nonumber\\
& & -\left[ (xt)^2-4 \right]{\sqrt{(xt)^2-1}\over xt} \nonumber\\
& & \times \left( 2\sqrt{1-t^2}-t^2\ln\left({1+\sqrt{1-t^2}\over 1-\sqrt{1-t^2}}\right) \right) \nonumber\\
& & + {1\over 2(xt)^3} 
      \Big[ \left( (xt)^2-16\right)\sqrt{(xt)^2-1} \nonumber\\
& & + 12xt\ln\left(xt+\sqrt{(xt)^2-1}\right)\Big] \nonumber\\ 
& & \times \Bigg[{2\over 3}(1-t^2)^{3/2} - t^2\ln\!\left({1+\sqrt{1-t^2}\over 1-\sqrt{1-t^2}}\right)\Bigg]
\end{eqnarray}
and the ellipticity parameter $\mu = [(\xi_1^2-\xi_2^2)/(\xi_1^2+\xi_2^2)]^2$.
This time the threshold behaviour ($\omega\approx m$) is found to be
\begin{eqnarray}
\label{Fmu}
\mathcal{F}(x,\mu) = {\pi\over 4}\mu (x-1)^2. 
\end{eqnarray}
Hence, in the general case, the cross section for two-photon pair creation close to threshold
raises like the second power of the excess energy $\omega-m$. Only in the special case
of circular polarization ($\mu=0$) the threshold cross section scales with the fourth
power as given in Eq.\,(\ref{2phlimit}). We notice, however, that our first-order treatment
of the Coulomb field requires the parameter $Z\alpha/v_\pm$ to be small, where $v_\pm$ 
denote the velocities of the created particles. For pair production very close to the 
threshold, Coulomb corrections are expected to be important.
A summary of our results on two-photon pair production for linear and circular polarization 
including various approximations is displayed in Fig.\,3.

\begin{figure}[ht]
\begin{center}
\includegraphics[width=6cm]{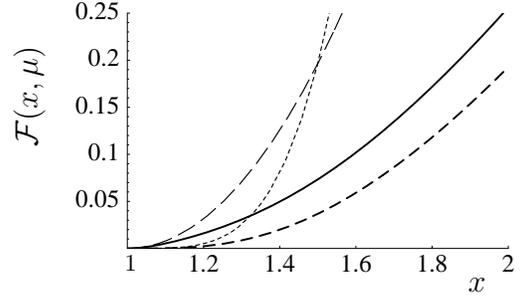}
\begin{picture}(0,0)(0,0)
\put(-23,-5){\large$x$}
\put(-153,3){${\rm 1}$}
\put(-195,50){\rotatebox{90}{\large$\mathcal{F}(x,\mu)$}}
\end{picture}
\caption{\label{Fcirclin} The function $\mathcal{F}(x,\mu)$ in the intervall $1\le x\le 2$
for $\mu=1$ (linear polarization, solid line) and $\mu=0$ (circular polarization, 
short-dashed line). The long-dashed and dotted lines show the respective threshold laws for 
$x\approx 1$ according to Eqs.\,(\ref{Flim}) and (\ref{Fmu}).} 
\end{center} 
\end{figure}

When applying our general formula (\ref{sigmagen}) to the same parameters as above but for a 
linearly polarized laser wave 
($\mu=1$, $\xi_1/\sqrt{2}=7.5\times 10^{-4}$, $\xi_2=0$, $\omega = 900$ keV, and $Z=1$), we get
a total production rate of 290 sec$^{-1}$, which is in rather good agreement with the corresponding
value of 240 sec$^{-1}$ found in Ref. \cite{MVGff2}. The difference is most likely due to small 
inaccuracies in the 5-dimensional integration along with the numerical evaluation of the generalized
Bessel functions in Ref. \cite{MVGff2}. As is seen from Fig.\,3, the pair production rate for 
a linearly polarized laser wave is noticeably larger than the 
corresponding rate for circular laser polarization within the whole frequency range $m <\omega< 2m$.

\subsection{High-frequency limit}
In this subsection we analyse the high-energy behaviour of strong-field pair creation,
i.e., we shall assume $\omega\gg m$ but allow for arbitrary value of $\xi$.
Let us first consider the circular polarization case and start from Eq.\,(\ref{circ2}). 
We exchange the order of 
integration and first integrate over $t$. Since $\omega\gg m$, we can introduce a
splitting parameter $\epsilon$ with $a\ll\epsilon\ll 1$, and divide the $t$-integration 
into ranges from $0$ to $\epsilon$ and from $\epsilon$ to 1. In the first region we 
have $t\ll 1$. Substituting $s=1/t$ we get
\begin{eqnarray}
\int_0^\epsilon dt\, \{...\} &=& \int_{1/\epsilon}^\infty ds\,\bigg\{
{8\xi^2\sin^2\rho\over 3z}Y_1(2a\rho z s) \nonumber\\
& & + {2\over 3}\left[ Y_1(2a\rho s) - {1\over z}Y_1(2a\rho z s) \right] \bigg\}.
\end{eqnarray}
With the help of the identity $Y_1(x) = -{d\over dx}Y_0(x)$ this is readily integrated to give
\begin{eqnarray}
\int_0^\epsilon dt\, \{...\} &=& {2\xi^2\over 3\pi a\rho z^2} \bigg\{
\left[ 4\sin^2\rho + \left(1-{\sin^2\rho\over\rho^2}\right) \right] \nonumber\\
& &\!\!\!\!\!\!\!\!\!\!\!\!\!\!\! \times \left( \ln{a\rho\over\epsilon} + C \right)
+\left( 4\sin^2\rho - {1\over\xi^2} \right)\ln z \bigg\}
\end{eqnarray}
where the relation $Y_0(x)\approx (2/\pi)[\ln(x/2)+C]$ was used, 
which is valid for $x\ll 1$; $C$ denotes Euler's constant.
In the second range we have $a\ll t$ and we can exploit the small-argument behaviour
of the Neumann function: $Y_1(x)\approx -2/(\pi x)$. Then the integration is easily performed:
\begin{eqnarray}
\int_\epsilon^1 dt\, \{...\} &=& -{2\xi^2\over 3\pi a\rho z^2}\bigg\{ \ln{2\over\epsilon}
\left[ 4\sin^2\rho + \left(1-{\sin^2\rho\over\rho^2}\right) \right] \nonumber\\
& & - {19\over 3}\sin^2\rho - {4\over 3}\left(1-{\sin^2\rho\over\rho^2}\right) \bigg\}.
\end{eqnarray}
Putting both parts together, $\epsilon$ drops out as it should and we arrive at
\begin{eqnarray}
\label{WHE}
\dot W &=& {2\over 3\pi^2}(Z\alpha)^2{m\xi^2\over a} \int_0^\infty {d\rho\over\rho^2 z^2}
\nonumber\\
& & \times\bigg\{ \left[ 4\sin^2\rho + \left(1-{\sin^2\rho\over\rho^2}\right) \right]
\left( \ln{2\over a\rho z} - C \right) \nonumber\\
& & + {z^2\over\xi^2}\ln z
- {19\over 3}\sin^2\rho - {4\over 3}\left(1-{\sin^2\rho\over\rho^2}\right) \bigg\}.
\end{eqnarray}
Applying an integration by parts to one of the terms, this can be rewritten as
\begin{eqnarray}
\label{rho}
\dot W &=& {2\over 3\pi^2}(Z\alpha)^2{\xi^2\over a}m \int_0^\infty {d\rho\over\rho^2 z^2}
\nonumber\\
& & \times\bigg\{ \left[ 4\sin^2\rho + \left(1-{\sin^2\rho\over\rho^2}\right) \right]
\left( \ln{2\over a\rho z} - C \right) \nonumber\\
& & + \left( 1-{\sin 2\rho\over 2\rho}\right)
- {19\over 3}\sin^2\rho - {7\over 3}\left(1-{\sin^2\rho\over\rho^2}\right) \bigg\}.\nonumber\\
\
\end{eqnarray}
At $\xi\ll 1$, the cross section resulting from Eq.\,(\ref{rho}) in the leading order in an 
expansion with respect to $\xi^2$ coincides with Eq.\,(\ref{BH}). The next-to-leading order
term for the cross section reads 
\begin{eqnarray}
\label{delta}
\Delta\sigma = -{52\over 45}Z^2\alpha\xi^2 r_e^2
               \left[ \ln\left({2\omega\over m}\right) - {22\over 13}\ln 2 - {124\over 195}\right].
\end{eqnarray}
The correction $\Delta\sigma$ is negative. This is due to the fact that it not only 
contains the process of two-photon pair production but also includes the $\xi^2$-corrections
to one-photon pair creation (cf. Fig.\,1). The latter is negative and has a larger absolute 
value than the contribution from two-photon pair creation. 

Now we consider the case $\xi\gg 1$, which corresponds to a supercritical laser field 
strength $E\gg E_c$. Note that, for the very high frequencies ($\omega\gg m$) assumed,  
such extremely large field strengths are far beyond present and near-future technical 
capabilities.
In this situation, the main contribution to the integral over $\rho$ in Eq.\,(\ref{rho}) 
comes from the region $\rho\sim 1/\xi\ll 1$. Performing the according 
expansion of the integrand and taking the integral, we obtain
\begin{eqnarray}
\label{sigmaHE}
\sigma = {26\over 3\sqrt{3}}{Z^2\alpha\over\xi}r_e^2
\left[ \ln\left({\omega\xi\over 2\sqrt{3}m}\right) - C - {58\over 39} \right].
\end{eqnarray}
We notice that the leading logarithm in Eq.\,(\ref{sigmaHE}) has already been derived
by Yakovlev \cite{Yak}. Further, Eq.\,(\ref{sigmaHE}) agrees with the corresponding
asymptotic limits found in Ref.\,\cite{NN} for pair creation by a Coulomb field 
and a constant crossed field and in Ref.\,\cite{BKS2} for pair creation in a Coulomb
field and a constant homogeneous field. It is worth stressing that the $\xi$-dependence in Eq.\,(\ref{sigmaHE}) cannot simply be expressed in terms of the effective (laser-dressed) 
electron mass $m_\ast = m\sqrt{1+\xi^2}$ \cite{LL}.

For the general case of an elliptically polarized laser wave the expression
corresponding to Eq.\,(\ref{WHE}) reads
\begin{eqnarray}
\dot W &=& {(Z\alpha)^2\omega\over 3\pi^2} \int_0^\infty {d\rho\over\rho^2}
\Bigg\{ \left[ \ln\left({4\over a\rho\chi}\right) - C\right] \nonumber\\
& & \times \Bigg[{\xi_1^2+\xi_2^2\over g_2}(A_1+2\sin^2\rho) - 2\ln{\chi\over 2} \nonumber\\
& & +{1\over A_0}\left({g_0\over g_2}-1\right)(A_1-3\sin^2\rho)\Bigg] \nonumber\\
& & +{\xi_1^2+\xi_2^2\over g_2}
   \Bigg[A_1\left(\ln{g_0+g_2\over 2g_2} - {4\over 3}\right) \nonumber\\
& & +2\sin^2\rho\left( \ln{g_0+g_2\over 2g_2} - {19\over 12} \right)\Bigg] \nonumber\\
& & -\ln{\chi\over 2}\left(\ln{\chi\over 2}-{5\over 3}\right)
-{1\over 2}L\left({g_2-g_0\over g_2+g_0}\right) \nonumber\\
& & +{1\over A_0 g_2}\Bigg[ A_1\left(g_0\ln{g_0+g_2\over 2g_2}-{4\over 3}(g_0-g_2)\right)
\nonumber\\
& & -3\sin^2\rho\left(g_0\ln{g_0+g_2\over 2g_2}-{3\over 2}(g_0-g_2)\right)\Bigg]\Bigg\}
\end{eqnarray}
with $g_2 = \sqrt{g_0^2-g_1^2}$, $\chi = \sqrt{2(g_0+g_2)}$, and Spencer's function
\begin{eqnarray}
L(x) = \int_0^x{dy\over y}\ln(1+y).
\end{eqnarray}
In the limit $(\xi_1^2+\xi_2^2)^{1/2}\gg 1$, we again can expand the integrand for small
values of $\rho$ and, within logarithmic accuracy, arrive at the following cross section:
\begin{eqnarray}
\label{sigmaHEgen}
\sigma &=& {26\over 3\sqrt{3}} {Z^2\alpha\over\sqrt{{1\over 2}(\xi_1^2+\xi_2^2)}}\ r_e^2\
_2F_1\!\left(-{1\over 4},{1\over 4},1,\mu\right) \nonumber\\
& & \times \ln\!\left({\omega\over m}\sqrt{\xi_1^2+\xi_2^2}\right).
\end{eqnarray}
Here, $_2F_1$ denotes a hypergeometric function, which smoothly depends on the ellipticity 
parameter $\mu$ and monotonously decreases from unity for $\mu=0$ (circular polarization) 
to $2\sqrt{2}/\pi\approx 0.9$ for $\mu=1$ (linear polarization).

\subsection{Quasiclassical limit}
Now we consider the limit $\xi\gg 1$. Note that today's most powerful lasers achieve
$\xi\sim 10^2$. In this limit, the main contribution to the integral over $\rho$ in
Eqs.\,(\ref{circ2}) and (\ref{gen2}) comes again from the region $\rho\sim 1/\xi\ll 1$. 
Hence, for the case of circular polarization, we can write
\begin{eqnarray}
\label{circquasi}
\dot W &=& -{(Z\alpha)^2\over\pi}m \int_0^1 dt \int_0^\infty {d\rho\over\rho} \bigg\{
{6\rho^2\over u} Y_1(\beta u) \nonumber\\
& & \times\left[{2\over 3}(2+t^2){\sqrt{1-t^2}\over t^2}-
\ln\left({1+\sqrt{1-t^2}\over 1-\sqrt{1-t^2}}\right)\right] \nonumber\\
& & + {2\over 3}\left[ Y_1(\beta)-{1\over u}Y_1(\beta u)\right]{(1-t^2)^{3/2}\over t^2} \nonumber\\
& & + \Big[ Y_1(\beta)- u Y_1(\beta u)\Big]
\ln\left({1+\sqrt{1-t^2}\over 1-\sqrt{1-t^2}}\right) \bigg\},
\end{eqnarray}
where now $\beta = 4\sqrt{3}\rho/(t\eta)$ with $\eta = \omega\xi/m = E/E_c$ and 
$u = \sqrt{1+\rho^2}$. Concerning the laser field parameters, the total rate in 
Eq.\,(\ref{circquasi}) only depends on the dimensionless ratio $\eta$, which is the 
so-called quasiclassical parameter. In the limit $\eta\gg 1$ we arrive at 
the production rate
\begin{eqnarray}
\label{rateloweta}
\dot W = {13\over 6\sqrt{3}\pi}(Z\alpha)^2 m\,\eta
\left[ \ln\left({\eta\over 2\sqrt{3}}\right) - C - {58\over 39} \right],
\end{eqnarray}
which corresponds to the cross section (\ref{sigmaHE}) obtained in Sec.\,III.C. 
In the opposite limit $\eta\ll 1$, which corresponds to the 
existing high-power lasers in the optical or infrared frequency range, the integral in 
Eq.\,(\ref{circquasi}) can be evaluated using Laplace's method. The resulting 
pair creation rate
\begin{eqnarray}
\label{sigmaquasi}
\dot W = {(Z\alpha)^2\over 2\sqrt{\pi}}\, m \left({\eta\over 2\sqrt{3}}\right)^{5/2}
{\rm exp}\!\left(-{2\sqrt{3}\over\eta}\right)
\end{eqnarray}
exhibits a tunneling behaviour. Equation\,(\ref{sigmaquasi}) is in agreement with
corresponding results on pair production in a Coulomb field and a constant crossed 
field \cite{NN} or a Coulomb field and a constant homegenous field \cite{BKS2}.

In a similar way we can derive from Eq.\,(\ref{gen2}) the $\eta\ll 1$ limit for an 
elliptically polarized laser wave. To this end, we assume $\eta_1 > \eta_2$ and $(\eta_1^2-\eta_2^2)/\eta_1^2\gg\eta_1$, where $\eta_j = \omega\xi_j/m$ for $j=1,2$.
The latter condition means, that the laser field is considerably different from a 
circularly polarized wave. Then we find
\begin{eqnarray}
\label{ratebigeta}
\dot W = {(Z\alpha)^2\over\sqrt{2}\pi}m\sqrt{\eta_1^2\over \eta_1^2-\eta_2^2}\,
\left({\eta_1\over 2\sqrt{3}}\right)^3 {\rm exp}\!\left(-{2\sqrt{3}\over\eta_1}\right).
\end{eqnarray}
In comparison with the result for circular polarization in Eq.\,(\ref{sigmaquasi}),
the production rate in Eq.\,(\ref{ratebigeta}) is suppressed by an additional factor
of $\sqrt{\eta_1}$. This suppression has the same nature as in the case of $e^+e^-$
pair production by a real photon in a laser field \cite{NR}. The reason for the
suppression is that, in general, the modulus of the field strength is truely oscillating, 
while in a circularly polarized wave it has a constant value. We note that Eq.\,(\ref{ratebigeta}) can also be derived by averaging Eq.\,(\ref{sigmaquasi}) over one oscillation cycle.

\section{Conclusions}
We have investigated in detail the total probability for electron-positron pair creation by a 
nuclear Coulomb field and an intense laser field. Employing the optical theorem in connection 
with the known polarization operator of a photon in an electromagnetic plane wave, we have 
derived compact, general expressions for the total pair production rate and found explicit 
formulae for various intensity and frequency regimes of interest. We have demonstrated that 
significant differences exist between the general case of an elliptically polarized laser 
wave and the special case of circular polarization [see, e.g., Eqs.\,(\ref{Fmu}) and
(\ref{ratebigeta}) versus Eqs.\,(\ref{Flim}) and (\ref{sigmaquasi})]. Particular emphasize 
was placed on the nonlinear process of two-photon pair creation. 

Let us estimate the feasibility of $e^+e^-$ pair production using an XFEL beam 
\cite{XFEL,ringwald} and 
a beam of relativistic protons. We will compare the following possible regimes of interaction: 
1) the one-photon perturbative regime, where $\xi\ll 1$ and $\omega \gg m$ [see Eq.\,(\ref{BH})], 
2) the two-photon perturbative regime, where $\xi\ll 1$ and $m < \omega < 2m$  
[see Eq.\,(\ref{sigmagen})], 3) the tunneling regime, where $\xi\gg 1$  and $\xi \ll m/\omega$ 
[see Eq.\,(\ref{sigmaquasi})], and 4) the overcritical-field regime, where $\xi\gg 1$ and 
$\xi \gg m/\omega$ [see Eq.\,(\ref{rateloweta})]. We have to take into account that 
the mentioned formulae are valid in the rest frame of the proton, which moves with a large
Lorentz-factor $\gamma$. In the rest frame of the proton, the photon frequency is increased with respect to its value in the lab frame: $\omega\approx 2\gamma \omega_L$, while the XFEL pulse duration is decreased: $\tau\approx\tau_L/2\gamma$. Here, $\omega_L$ and $\tau_L$ denote the laser 
frequency and pulse duration in the lab frame, respectively. Note that, nowadays, the proton 
beam at DESY can be accelerated up to 920 GeV, which corresponds to $\gamma\approx 1000$.

For the following estimates, we will throughout assume an XFEL pulse duration of $\tau_L=100$ fs.
The XFEL beam collides with a bunch of protons moving at $\gamma=1000$, containing 10$^{11}$ particles, and having a beam radius of 30 $\mu$m.
In the one-photon perturbative regime, assuming an XFEL photon energy of $\omega_L=10$ keV, 
an intensity parameter of $\xi=10^{-5}$, and a beam radius of $R=30$ $\mu$m, we find that
$\sim 100$ pairs are produced per collision. For a given laser intensity, the number of
created particles is independent of the laser's polarization state. 
In the second regime of two-photon pair creation, employing a linearly polarized XFEL beam 
is more favorable. Supposing the collision parameters $\omega_L=0.5$ keV, $\xi_1=3\times 10^{-3}$,
and $R=30$ $\mu$m, the process proceeds far above threshold and one pair can be produced per 
shot. 
In the overcritical-field regime, circular polarization of the XFEL beam leads to higher
production yields. In this regime, a large $\xi$-value is necessary, which can be achieved
by focusing the XFEL beam \cite{ringwald}. For $\xi=3$, $\omega_L=10$ keV and $R=0.4$ nm, 
one pair can be produced per collision. Here one has to take into account that the XFEL
pulse will interact only with a very small fraction of the proton bunch. For the assumed
parameters, the latter has a length of 10 cm and contains a particle density of 10$^{15}$ cm$^{-3}$. 

Thus, we see that, in principle, experimental studies of the pair production process are 
possible by using an XFEL and a relativistic proton beam. Employing a non-focused XFEL beam 
(cf. the so-called design and available parameters in Table 1 of Ref.\,\cite{ringwald}) and 
an ultra-relativistic proton beam, the perturbative regimes of pair creation can be realized, including nonlinear pair production via two-photon absorption. The realization of the 
strong-field regimes with $\xi\gg 1$ will require focusing of the XFEL beam (cf. the goal 
parameters in Table 1 of Ref.\,\cite{ringwald}). In this case the interaction volume, i.e. 
the region of beam overlap, is extremely small which substantially decreases the total 
production probability. As a result, the pair creation seems unrealistic in the tunneling 
regime but, at least in principle, feasible in the overcritical-field regime.

In summary, with the novel sources of spatially coherent x-rays along with the relativistic 
proton beams of DESY, the results on pair production presented in this paper could be tested.

\section*{Acknowledgements}
A.I.M. thanks the Max-Planck-Institute for Nuclear Physics for its kind hospitality
during an extended guest researcher appointment when this work was done.
U.D.J. acknowledges support from the Deutsche Forschungsgemeinschaft (Heisenberg program).

\end{document}